\documentstyle[aps]{revtex}
\begin{document}
\draft
\title{High frequency vortex dynamics and magnetoconductivity
  of high temperature superconductors}
\author{Jan Kol\'a\v{c}ek}
\address{Institute of Physics, ASCR, Cukrovarnick\'a 10,16200 Prague 6,
Czech Republic}
\author {Etsuo Kawate}
\address {National Research Laboratory of Metrology,
1-1-4, Umezono, Tsukuba, Ibaraki 305, Japan}
\date{\today}
\maketitle
\begin{abstract}
The vortex lattice with the superconducting and normal state charge carriers
fractions may be regarded as  three independent subsystems mutually
connected by interactions. The equations of motion for these three subsystems
must be solved simultaneously. In this way a new consistent theory of the
vortex dynamics is obtained and the magnetoconductivity calculated.
\end{abstract}

\section {INTRODUCTION}
The vortex dynamics attracts great attention from both the theoretical
and experimental point of view. Much controversy is concentrated on the
question, whether Magnus force is the only transverse force on the vortex
as is claimed by Ao and Thouless \cite{93Ao}, or if other ones, like
Iordanskii force from the normal fluid or Kopnin-Kraftsov force from
impurities may also contribute to the total transverse force. This topic
is discussed in detail by Sonin \cite{97Sonin}, where many relevant
references can be found.
\par
As the characteristic frequencies of the vortex system are in the far infrared
(FIR) region \cite{94Drew} the magnetooptical spectroscopy is a suitable tool
to study these problems. Recently results of the FIR magnetotransmission of
YBaCuO thin films using elliptically polarized light became available
\cite{96Lihn,96Wu}. To interpret the measurements, the high frequency
conductivity must be calculated. The first such calculation was made by
Gittleman-Rosenblum \cite{66Gittleman}. Later, the dissipation near the
flux line depinning threshold using the generalized temperature assisted
flux flow (TAFF) model was studied by Yeh \cite{91Yeh}. In these papers
the reaction of the superconducting (and normal) charge carriers on the
vortex movement was neglected. The redistribution of the induced ac
current density due to the vortex response is taken into account in the
Coffey-Clem \cite{91Coffey} calculation of the surface impedance.
Other references concerning this problem may be found in a recent
review \cite{96Golosovsky}.
\par
In principle, our approach is similar to the one of Coffey-Clem model,
but the formalism is different. We treat the  superconducting fluid,
normal fluid and the vortex system as three subsystems mutually connected
by interaction. Their equations of motion are solved simultaneously.
In this way a new, internally consistent theory of vortex motion is obtained.
We use this approach for calculating magnetoconductivity in high temperature
superconductor (HTSC) thin films. In this paper we consider the Magnus
force as the only interaction between the vortex system and superconducting
fluid and Lorentz force to be the main interaction force between the normal
fluid and vortex system in HTSC materials. It conforms to the necessary
symmetry requirements summarized by Sonin \cite{97Sonin} and has
the same sign and form as the D,D' terms derived by Stone \cite{96Stone} for
normal particles confined in the vortex core, if the limit of nonlocalized
particles is taken. We show, that including this transversal force from
normal state charge carriers gives reasonable magnetoconductivity tensor
which is compared with recent experimental data \cite{96Lihn}.

\section{THE  MODEL}
Let us consider the vortex lattice, in which the mean distance between
vortices is small in comparison with the penetration depth. In this case
the magnetic field in the superconductor is almost homogeneous. This model
may be appropriate for HTSC. In the field 5 T the mean distance between
vortices is about 20 nm, while in YBCO the London penetration depth
$\lambda_L$ is of the order 150 nm. The  diameter of the vortex core is
approximately equal to the coherence length $\xi$, which in YBCO is
about 2 nm. Therefore, it is possible to neglect possible redistribution
of the charge and current density in the vortex core and to take it into
account only as the source of vortex damping.
The damping force ${\bf F}_D$
is supposed to be proportional to the vortex velocity ${\bf v}_L$
with frequency independent viscosity coefficient $\eta$, so that
\begin{eqnarray} \label{eqdamping}
{\bf F}_{D} = -{\eta} {\bf v}_{L}
            = - {{m _{v}} \over {\tau _v}}  {\bf v}_L .
\end{eqnarray}
Introducing $m_v$ as the vortex mass per unit length, we will use the
vortex relaxation rate $1/\tau_v$ which is given in practical frequency
units, rather than the viscosity.
\par
For pinning we use the simplest model of a parabolic well, so that
the pinning force ${\bf F}_P$ is proportional to vortex displacement
${\bf r}_L$:
\begin{eqnarray} \label{eqpinning}
{\bf F}_{P} = -{\kappa} {\bf r}_{L}
            = - m _{v} {\alpha} ^{2}  {\bf r}_L
\end{eqnarray}
with $\kappa$ and $\alpha$ being the pinning constant and pinning frequency,
respectively.
\par
The interaction between superconducting fluid moving with velocity
${\bf v}_s$ and the vortex system is mediated by the Magnus force
\cite{93Ao} given by
\begin{eqnarray} \label{eqmagnusv}
{\bf F}_{M} (v) = {{n_s h} \over 2} ({\bf v}_s - {\bf v}_L ) \times {\bf z}
                = m _{v} f_s \Omega ({\bf v}_s - {\bf v}_L ) \times {\bf z},
\end{eqnarray}
where $n_s = f_s n$ is the density of superconducting fluid and ${\bf
F}_{M}(v)$ means, that this is the force felt by the vortex. The reaction
force ${\bf F}_{M} (s)$ acting on a superconducting particle is
\begin{eqnarray} \label{eqmagnuss}
{\bf F}_{M} (s) = - {{n_v} \over {n_s}} {\bf F}_{M} (v)
                = - m \omega _c ({\bf v}_s - {\bf v}_L ) \times {\bf z},
\end{eqnarray}
where $n_v$ is the vortex density (number of vortices per unit area),
$ \omega_c=eB/m = n_v h / 2m$ is the cyclotron frequency in the field
${\bf B} = n_v \Phi _0 {\bf z}$ caused by the vortex system and
$ \Phi_0$ is the magnetic flux quantum.
In (\ref {eqmagnusv}) we introduced the frequency of the cyclotron vortex
motion $ \Omega = {{n h} / {2 m_v}} $. It is interesting to note that,
using the Hsu's expression for the vortex mass
$ m _{v} = {{(\pi}^2} /4) m k_{F}^{2} {\xi}^{2} $
\cite {93Hsu}, the 2D expression for the Fermi wave vector
$ k_{F}^{2} = 2 \pi n $ and
$ \xi = {{\hbar v_F} / {\pi \Delta}}$ for the coherence length,
it is possible to show, that
$\Omega = \Delta^2/E_F$ ($\Delta$ is the gap and $E_F$ is the Fermi energy),
which is the level separation in the vortex core \cite {96Golosovsky}.
\par
The interaction between the vortex system and the normal state fluid may
be obtained in the following way. From the Aharonov-Casher Lagrangian 
\cite {84Aharonov} it can be shown that if the vortex lattice moves with
velocity ${\bf v}_L$, the force imposed by the vortex system on one 
normal state particle moving with velocity ${\bf v}_n$ is 
\begin{eqnarray} \label{eqLorentzn}
{\bf F}_{L} (n) = {{n_v h} \over {2}} ({\bf v}_n - {\bf v}_L ) \times {\bf z}
                = m \omega _c ({\bf v}_n - {\bf v}_L ) \times {\bf z}.
\end{eqnarray}
According to the action-reaction law the vortex lattice
must feel the same force with opposite direction.
If there are $n_n=f_n n$  normal state particles,
the total force per unit length of one vortex is
\begin{eqnarray} \label{eqLorentzv}
{\bf F}_{L} (v) = - {{n_n} \over {n_v}} {\bf F}_{L} (n)
               = - f_n { {nh} \over 2} ({\bf v}_n - {\bf v}_L ) \times {\bf z}
               = - m_v f_n \Omega ({\bf v}_n - {\bf v}_L ) \times {\bf z}.
\end{eqnarray}
This expression is analogous to the Magnus force formula, but has
opposite sign. It satisfies the invariance requirements, 
according to which only the relative velocity of the particle 
with respect to the vortex system
is decisive. Factor $f_n$ is justified by the fact
that the total force is proportional to the number of particles
involved in the interaction
\par
The questions concerning forces acting on the vortex lattice are still 
seriously controversial. The Lorentz force (\ref{eqLorentzv}) following from
the Aharonov-Casher
Lagrangian is of electrodynamic origin, but similar formula is also used
to describe interaction of normal state fluid with vortices in neutral systems
(see e.g. \cite{97Sonin,96Stone,95Kopnin}).
Useful comments and replies regarding the spectral flow force and the Iordanskii 
force can be also found in 
\cite {98Ao_Kopnin,98Hall_Wexler,98Sonin_Wexler}.
Interaction of electric charge with moving vortex and the Aharonov-Casher
effect in two-dimensional superconductors was discussed e.g. by 
\v{S}im\'anek \cite{97Simanek}. Let us note  that it would
not be correct to consider the Lorentz force 
also for the superconducting fluid, which would
exactly cancel the Magnus force. The vortex lattice and the accompanying
magnetic field are created by superconducting current, so in this
case Lorentz force would mean "action on itself".
\par
Having draw up the interaction forces we will now write the equations
of motion for the three subsystems. As in the London model,
the superconducting fluid is supposed to move without damping,
\begin{eqnarray} \label{eqmotions}
m {\dot {\bf v}}_{s}  = e{\bf E} + {\bf F}_M (s) ,
\end{eqnarray}
while the normal state fluid motion is damped as in the conventional
Drude model
\begin{eqnarray} \label{eqmotionn}
m {\dot {\bf v}}_{n}  = e{\bf E} + {\bf F}_L (n) - {m \over \tau _n} {\bf v}_n
. \end{eqnarray}
Finally, for the vortex system we shall use the Newton type equation of
motion (of course the vortex mass and also all the forces are considered
per unit length)
\begin{eqnarray} \label{eqmotionv}
m_v {\dot {\bf v}}_{L}  = {\bf F}_P+{\bf F}_D + {\bf F}_M (v)+ {\bf F}_L (v)
. \end{eqnarray}

The system of three equations of motion (\ref{eqmotions}-\ref{eqmotionv})
together with expressions for the interaction forces
(\ref{eqmagnusv}-\ref{eqLorentzv}), damping and pinning force
(\ref {eqdamping},\ref{eqpinning}) form a closed set of equations for
the unknown ${\bf v}_L$, ${\bf v}_s$ and ${\bf v}_n$. Assuming a periodic
time dependence $e^{i \omega t}$, the three differential equations reduce
to the set of three linear equations :
\begin{eqnarray} \label{eqlinearsyst}
A_{ss} v_s +              A_{sv} v_L &=& {e \over m} E \nonumber \\
             A_{nn} v_n + A_{nv} v_L &=& {e \over m} E \\
A_{vs} v_s + A_{vn} v_n + A_{vv} v_L &=& 0             \nonumber
\end{eqnarray}
with the coefficients
\begin{eqnarray} \label{eqcoeff}
  A_{ss} &=& i(\omega - \omega _c )            \hspace{15mm};\hspace{5mm}
  A_{sv} =   i \omega _c  \nonumber \\
  A_{nn} &=& i(\omega + \omega _c - i/ \tau _n) \hspace{4mm};\hspace{5mm}
  A_{nv} =  -i \omega _c   \nonumber \\
  A_{vs} &=& i f_s \Omega                       \hspace{22mm};\hspace{5mm}
  A_{vn} = - i f_n \Omega \\
  A_{vv} &=& i(\omega + (f_n - f_s) \Omega - \alpha ^2/ \omega -
i/ \tau _v ) \nonumber
. \end{eqnarray}
If the determinant
$ D = A_{ss}A_{nn}A_{vv} - A_{nn}A_{sv}A_{vs} - A_{ss}A_{nv}A_{vn} $
is nonzero, the set can be readily solved to get
\begin{eqnarray} \label{eqveloc}
v_s \equiv g_s {eE \over m} &=& {{ A_{nn}A_{vv}
         + A_{sv}A_{vn} - A_{nv}A_{vn}} \over D} {eE \over m} \nonumber \\
v_n \equiv g_n {eE \over m} &=& {{ A_{ss}A_{vv}
         + A_{nv}A_{vs} - A_{sv}A_{vs}} \over D} {eE \over m} \\
v_L \equiv g_L {eE \over m} &=& {{-A_{ss}A_{vn}
          - A_{nn}A_{vs}               } \over D} {eE \over m} \nonumber
. \end{eqnarray}
Now it is straightforward to express the conductivity as
\begin{eqnarray} \label{eqcond}
\sigma = {j \over E} = {e \over E} (n_s v_s + n_n v_n)
         = \epsilon _0 \omega _p ^2 (f_s g_s + f_n g_n)
, \end{eqnarray}
where $ \omega _p = \sqrt{n e^2 / \epsilon _0 m} $
is the plasma frequency and the factors $g_s , g_n$ are defined
by eq. (\ref{eqveloc}).
\par
As expected, for physically meaningful parameters
$( \tau_v > 0, \tau_n > 0, \omega_c \Omega > 0 )$
the real part of conductivity is positive and the Kramers-Kronig
relation
$\sigma(\omega)=\sigma_0 + (\omega / i \pi)
  \int _{-\infty} ^{\infty} \!{\sigma (x) / (x^2 - x\omega )} {dx}$
as well as the f-sum rule
$ (1/ \pi) \int _{-\infty} ^{\infty}
  \!{{\it Re}(\sigma (\omega))}{d\omega} =
\epsilon_0 \omega_p^2 $ are satisfied.
The zero frequency limit of the conductivity
$\sigma_0 = \epsilon_0 \omega_p^2 \tau_n
(f_n + i[ \omega_c \tau_n (f_s - f_n) +
f_s/\omega_c \tau_n] )/(1+\tau_n^2 \omega_c^2)$
does not have the delta function component, as the pinning
constant is supposed to be finite, while pinning range
is infinite.
Expressing the conductivity tensor components as
$\sigma_{xx}(\omega) = (\sigma(\omega)+\sigma(-\omega))/2 $ ,
$\sigma_{xy}(\omega) = (\sigma(\omega)-\sigma(-\omega))/2 $,
it is possible to show, that also the Hall sum rule \cite{97Drew}
$ (1/ \pi) \int _{-\infty} ^{\infty}
\!{{\it Re}(t_H) }{d\omega} = \omega_H $, where
$t_H = \sigma_{xy}/\sigma_{xx} $ and
$ \omega_H = \lim_{\omega \rightarrow \infty} [-i \omega t_H (\omega)]
  = \omega_c(f_s -  f_n) $ is satisfied.
It is necessary to note, that without normal state fraction
( $f_n = 0$) the $t_H$ function has a pole at zero frequency, so
that in this case the Hall sum rule must be modified to
$ \omega_H = (\alpha^2 \omega_c /( \Omega \omega_c + \alpha^2))
+ (1 / \pi) \int _{-\infty} ^{\infty}
\!{{\it Re}(t_H) }{d\omega} $.

\section {Absence of normal state fluid}
Usually it is considered that at zero temperature all the charge carriers
condense, so that normal state fluid is absent. It is not necessary true
for all materials, but it is useful to discuss this limit first.
\par
For free vortices (vortices without pinning and damping) the two equations
of motion
$m {\dot {\bf v}}_{s}  =  - m \omega _c ({\bf v}_s - {\bf v}_L ) \times {\bf z} $
for the superconducting fluid and
$m {\dot {\bf v}}_{L}  =  m _{v} f_s \Omega ({\bf v}_s - {\bf v}_L ) \times {\bf z} $
for the vortex system are readily simplified to
$v_L / v_s = \Omega/(\Omega-\omega)$ and
$v_L / v_s = (\omega_c - \omega)/\omega_c$, respectively.
Consequently, two nontrivial solutions exist: for zero frequency
$v_L=v_s$, while for  $\omega=\Omega+\omega_c$ the velocity ratio is
$v_L / v_s =-\Omega/\omega_c$. This means that the superconducting liquid
and vortices may move either in parallel with constant velocity (this
solution is required by Galilean invariance), or may oscillate with
opposite phase, with the inertial center remaining at rest.
\par
In general, with $f_n =0$ the coefficient $A_{vn}$ equals zero and the
conductivity formula (\ref {eqcond}) reduces to
\begin{eqnarray} \label{eqcondzfn}
\sigma(f_n=0) = \epsilon _0 \omega _p ^2 {A_{vv} \over
                   {A_{ss} A_{vv} - A_{sv} A_{vs}} }
. \end{eqnarray}
Let us note that in the limit of zero vortex density
($\omega_c \rightarrow 0 $), this formula reduces to the London expression
for conductivity $\sigma = \epsilon _0 \omega _p ^2 / i \omega $,
as expected. For zero pinning (but nonzero damping) the explicit expression
for conductivity may be written as:
\begin{eqnarray} \label{eqcondzp}
\sigma(f_n=0,\alpha=0)
  =  { {1+i\tau_v(\omega - \Omega)} \over
     { \tau_v \omega(\omega_c + \Omega - \omega) + i(\omega - \omega_c)} }
. \end{eqnarray}
It is clear that in this case the real part of conductivity is nonzero
even at zero frequency
$\sigma_1(f_n=0,\alpha=0,\omega=0)=\epsilon_0 \omega_p^2 \tau_v
\Omega / \omega_c $.
Contrary to it, for nonzero pinning we get
$\sigma(f_n=0,\alpha\not=0,\omega=0)=\epsilon_0 \omega_p^2 i/\omega_c $
with zero real part of conductivity. This result is understandable,
if we recall that in our simple model the pinning barrier
is infinite, so that the d.c. transport must be nondissipative.
\par
In reality the pinning barrier is not infinite. Depending on frequency,
temperature, magnetic field, as well as density and strength of pinning
sites, various regimes as flux creep, flux flow, temperature assisted
flux flow etc. \cite {94Blatter} can be recognized. To keep the discussion
simple, we will analyze just two simple limits. In the "full pinning" (FP)
limit the driving field is low, so that each vortex is bound to the individual
pinning valley, making only small oscillations. In this case the pinning
force plays an important role. Contrary to it in the limit of high driving
field the amplitude of the vortex oscillation is larger than the distance
between the pinning centers, and the averaged pinning force is effectively
zero (ZP). In the intermediate state the pining force is nonzero, but not
proportional to the distance from the pinning center which leads to
nonlinear effects. We will show that, in some frequency range, nonlinear
effects can be expected even at relatively low fields which are commonly
used in laboratory experiments.
\par
Let us estimate the realistic values for the parameters of the theory.
For coherence length $\xi=2 \text{ nm}$, and effective mass $m=4m_e$ using
the Hsu's expression for the vortex mass \cite{93Hsu}, we can
estimate $\Omega=2\hbar/(\pi^2 m\xi^2)=49 \text{ cm}^{-1}$. The cyclotron
frequency in the field  4T is $5.9 \text{ cm}^{-1}$.
Using the expression  \cite{96Golosovsky}
$\kappa = (0.01 \div 0.05) \mu_0 H^2_c$
for the pinning coefficient, and the vortex mass estimation \cite{91Yeh}
$m_v=1.6*10^{10} m_e/m$, the range for pinning frequency
$\alpha=19 \div 95 \text{ cm}^{-1}$ may be obtained. As
we did not select any model for the vortex damping, we leave this parameter
as free. To make a model calculation we used the following set of parameters:
$\omega_c=5$, $\Omega=50$, $ \alpha=30$, $\omega_p=6000$, $1/\tau_v=10$
(all values are in $\text{cm}^{-1}$). The conductivity calculated
in FP and ZP limits are displayed in fig.1. The conductivity peaks are
expected near the frequencies, where the real or imaginary part of
the determinant $D=A_{ss}A_{vv}-A_{sv}A_{vs}$ which appears in the
denominator of (\ref {eqcondzfn}) is zero.
In ZP limit the expected peak values of conductivity are
\begin{eqnarray} \label{eqpeaks}
\sigma_1(\omega=0)               &=&
            \epsilon_0 \omega_p^2 \tau_v \Omega / \omega_c \nonumber \\
\sigma_1(\omega=\Omega+\omega_c) &=&
             \epsilon_0 \omega_p^2 \tau_v \omega_c / \Omega           \\
\sigma_1(\omega=\omega_c)        &=&
              \epsilon_0 \omega_p^2/\tau_v \Omega   \omega_c \nonumber
. \end{eqnarray}

In fig.1. only two sharp peaks are present for the ZP limit (dashed line).
It is obvious from (\ref{eqpeaks}) that, while for low vortex damping
(large $\tau_v $) the peaks at eigenfrequencies of the system
(0 and $\Omega+\omega_c$ ) are important, for high vortex damping
the peak at $\omega_c$ will dominate. This is illustrated in fig.2,
where the conductivity for vortex damping $1/\tau_v$  from 10 to 200 are
displayed. It is possible to see, how with increasing vortex damping
the peak shifts from zero frequency to the cyclotron frequency
$\omega_c$. For FP limit, due to the pinning term $\alpha^2/\omega$ the order
of the determinant D is higher in $\omega$, so one more peak is expected
in accord with the model calculation results displayed in fig.1 (solid line).
\par
In fig.3 the relative value of the vortex oscillation amplitude
$a_v = |{\bf r}_L| m / eE $ is shown as a function of $\omega$. It is
clear that, while at high  frequency  the oscillation amplitude is low
so that FP limit is appropriate, at lower frequencies the amplitude is high,
so that ZP  limit must be adopted. In principle, beside the pinning
frequency $\alpha$ determining the pinning force at low
oscillation amplitude, two characteristic lengths $r_1$ and $r_2$,
the amplitudes of vortex oscillation, at which the pinning force declines
from the linear law and at which the pinning force is effectively zero,
should be introduced. In this way, for a given driving field E,
two crossover frequencies $\omega_{d_1,d_2} = \sqrt {eE/r_{1,2} m} $
are defined. For $ a_v<1/\omega_{d_1}^2$   the FP limit is valid,
while for $a_v>1/\omega_{d_2}^2$  the ZP must be used. If neither
condition is fulfilled, the system is in a nonlinear region, where
the conductivity depends on the driving field strength. It is
interesting to note, that for some frequencies both conditions
$ a_v(FP)<1/\omega_{d1}^2$ and $a_v(ZP)>1/\omega_{d2}^2$ may be fulfilled
at one time. This means, that in this frequency region  bistability
may occur. Depending on the history, at the same experimental conditions
two regimes - the low and high vortex oscillation amplitude, corresponding
to the low and high resistivity state - may be achieved! All these
possibilities are illustrated in fig.3. If we estimate the range of
pinning force ($r_d$) to be about $10 \text{ nm}$ and if the intensity of
radiation used for measurement is $1 \text{ mW/mm}^2$ so that the driving
field E is of order $1.7 * 10^4 \text{ V/m}$,
we get $\omega_d= 9 \text{ cm}^{-1}$. For illustration purposes we have
chosen $\omega_{d_1}= \omega_{d_2}=10 \text{ cm}^{-1}$.
It is obvious that, depending on the frequency and intensity of the radiation
used for the measurements, many interesting nonlinear effects may be expected.

\section{Influence of normal state fluid}
For nonzero temperatures there are two contributions to the real conductivity.
One is connected with the normal state charge carriers, the other with
vortices. As expected, without vortices we get
$\sigma(\omega_c = 0) = \epsilon_0 \omega_p^2 [f_s/i\omega + f_n\tau_n/(1+i\omega\tau_n)]$ ,
which is the sum of the London and Drude model contributions.
The normal state limit  $(f_s \rightarrow 0 )$  does not have much sense,
as without superconducting fraction we can not have any vortices.
However, if we simulate external magnetic field by making vortices
unable to move, we should get the formula for a normal conductor in
magnetic field. Indeed, in the limit of vortices fixed to the lattice
($\alpha\rightarrow\infty$) or of infinite vortex mass ($\Omega=0$),
we get the expected result
$\sigma(f_s=0,\Omega=0)=\sigma(f_s=0,\alpha\rightarrow\infty) =
 \epsilon_0 \omega_p^2\tau_n /(1+i(\omega+\omega_c)\tau_n) $.
\par
The results of model calculations for $f_n=0.5$ in FP and ZP limits
are displayed in fig.4. To visualize the contribution of vortices,
the zero magnetic field conductivity ($\omega_c= 0$) is also
displayed in these graphs. In FP limit, when the amplitude of vortex
motion is small, almost all real part of conductivity originates from
the normal state charge carriers - except of the very
sharp feature near the zero frequency, which is caused by the vortex
resonance. On the other hand, in the ZP limit the conductivity is much
larger  and it is almost completely due to the vortex motion,
with the normal state fluid playing only a minor role. However, the
sharp vortex resonance peak is absent. It might be somewhat surprising,
that  the presence of vortices may slightly decrease the real part of
conductivity for some frequencies.
\par
It is instructive  to see, how increasing the normal state fraction
influences the conductivity. For the FP limit it is shown in fig.5.
We can see that with increasing $f_n$, the central peak
(connected with the normal state carriers conductivity) gradually
develops, while the side peaks diminish. Recently, Lihn at al.\cite{96Lihn}
measured far infrared magnetoconductivity tensor in YBaCuO
thin film. Their data are also displayed in fig.5 (dashed line).
The intensity of FIR radiation is usually rather small so the FP limit
could be appropriate. It is remarkable, that all the experimentally observed
features are quite well simulated by the curve with $f_n=0.3$.
This seems to indicate the presence of some normal state fraction
(probably located on CuO chains ) even at the lowest temperature.
Alternatively it may be due to the enhanced density of quasiparticles
in an applied magnetic field, as predicted for the d-wave superconductor
\cite {{93Volovik},{95Wang}}. It should be noted, that the sharp vortex
resonance peak on the theoretical curve is at lower frequency than
the range accessible by FIR spectroscopy, so it could not be observed
in the experiment.

\section {Conclusions}
Vortex lattice together with the superconducting and normal state fluid form three
subsystems mutually connected by interaction. Taking into account 
reaction forces by which vortices influence 
superconducting and normal state fluid and solving simultaneously
the three equations of motion a new, internally consistent
theory of vortex dynamics was developed. It was shown that due to the
finite range of the pinning force, at some frequencies nonlinear phenomena 
may be expected even for relatively low driving fields which are
commonly used in laboratory experiments. For comparison with experiment,
the knowledge of the power of radiation used for the measurements might
be crucial. The presented theory can qualitatively explain recent
measurements of far infrared magnetoconductivity tensor made by
Lihn at.al \cite{96Lihn}. The d.c. conductivity calculated in the framework 
of this model enables to explain theoretically controversial,
but experimentally firmly established Hall voltage sign reversal
\cite{Kolacek,98Kolacek}.

\acknowledgements
The authors are grateful to H.D.Drew and H.Lihn for providing
the original data from their magnetooptical measurement, as well as to
E.H.Brandt, E.\v{S}im\'anek and E.Sonin for helpful discussions.
This work was supported by grants
GA\v{C}R $\sharp202/96/0864$ and M\v{S}MT KONTAKT ME 160.
One of us (J.K.) thanks the Japanese International Superconductivity Center
(ISTEC) and New Energy and Industrial Technology Development Organization
(NEDO) for the fellowship, during which part of this work was done.



\begin{figure}
\caption{Real part of conductivity for FP (solid line) and ZP (dashed) limits. Parameters of the
model are $\omega_c=5$, $\Omega=50$, $\alpha=30$, $\omega_p=6000$,
$1/\tau_v=10$ (all in $\text{cm}^{-1}$).}
\label{fig1}
\end{figure}

\begin{figure}
\caption{Real part of conductivity for ZP limit, for $\omega_c=5, \Omega=50, \alpha=0,
\omega_p=6000$ and $1/\tau_v=10,20,50,100,200$ (all in $\text{cm}^{-1}$). For increasing vortex
damping the conductivity peak shifts from zero frequency to cyclotron frequency.}
\label{fig2}
\end{figure}

\begin{figure}
\caption{The relative vortex oscillation amplitude for FP (solid line) and ZP (dashed) limits
calculated using the same parameters as in Fig.1. The pinning range corresponding to crossover
frequency $\omega_d=10 \text{ cm}^{-1}$ is  marked by a horizontal line. The frequency regions marked by
$\backslash \backslash \backslash $ (///) are
regions where FP (ZP) limit are appropriate.}
\label{fig3}
\end{figure}

\begin{figure}
\caption{The comparison of real part of conductivity in the FP (solid line) and ZP (dashed) limits
with conductivity in zero magnetic field (dotted) in presence of normal state fraction $f_n=0.5$;
$1/\tau_n=15$.  All the other parameters are same as in the Fig.1. }
\label{fig4}
\end{figure}

\begin{figure}
\caption{Real part of conductivity in FP limit
for normal state fraction $f_n=$ 0.1, 0.15, 0.2, 0.25, 0.3
(the tendencies for increasing $f_n$ are marked by arrows).
Other parameters are same as in Fig.1. The
experimental data obtained by Lihn at al. [4] are also plotted
(dashed line) for comparison.}
\label{fig5}
\end{figure}
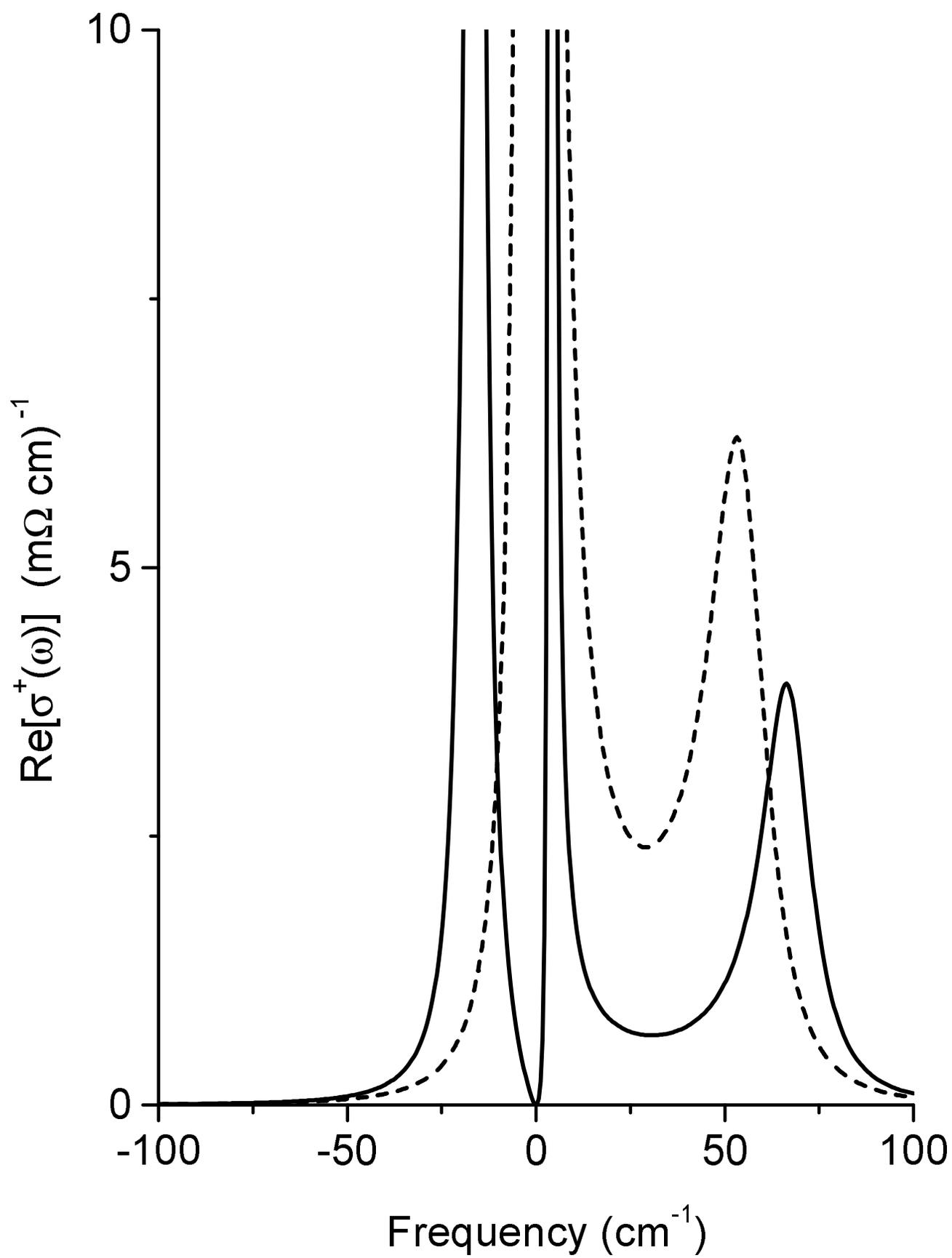
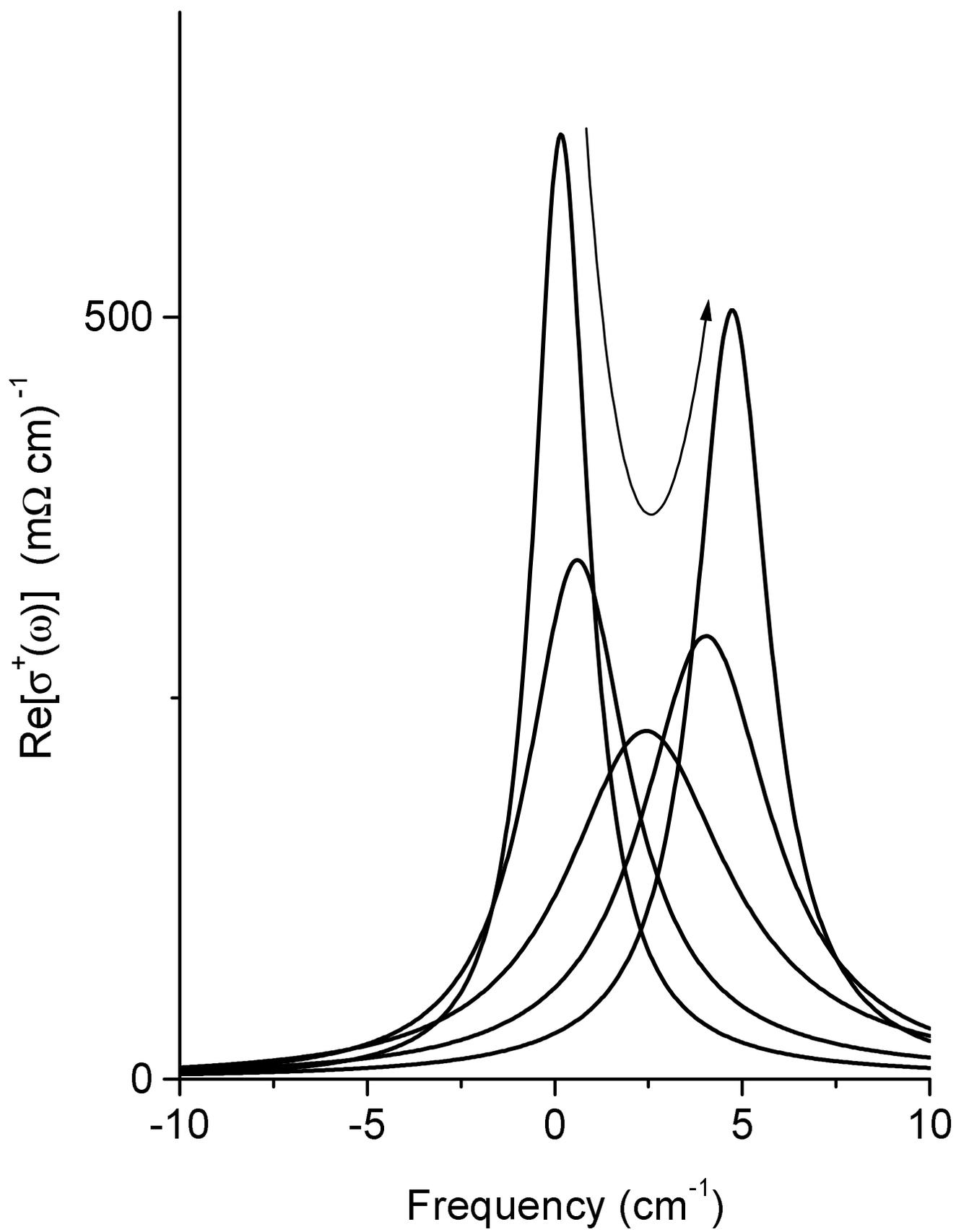
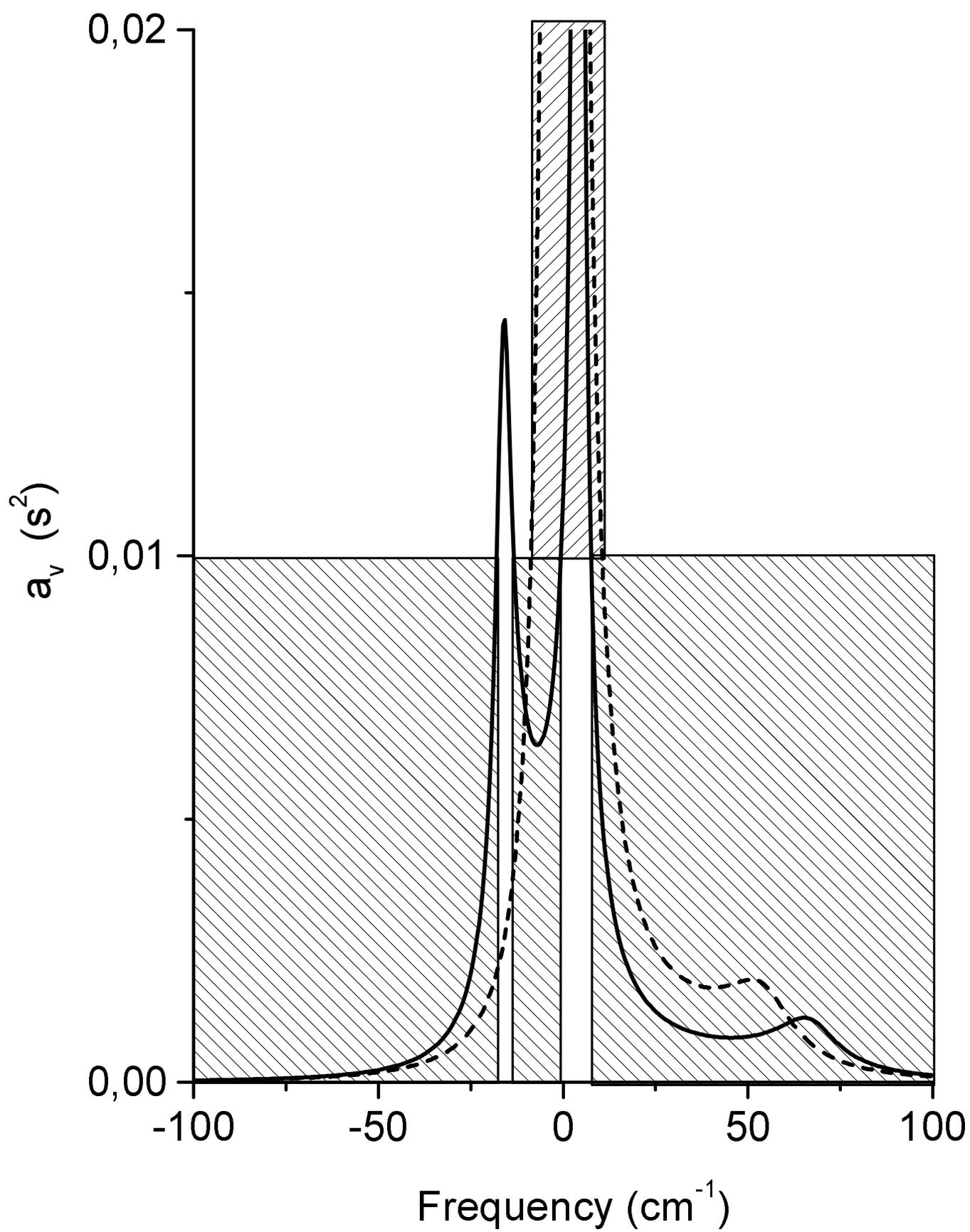
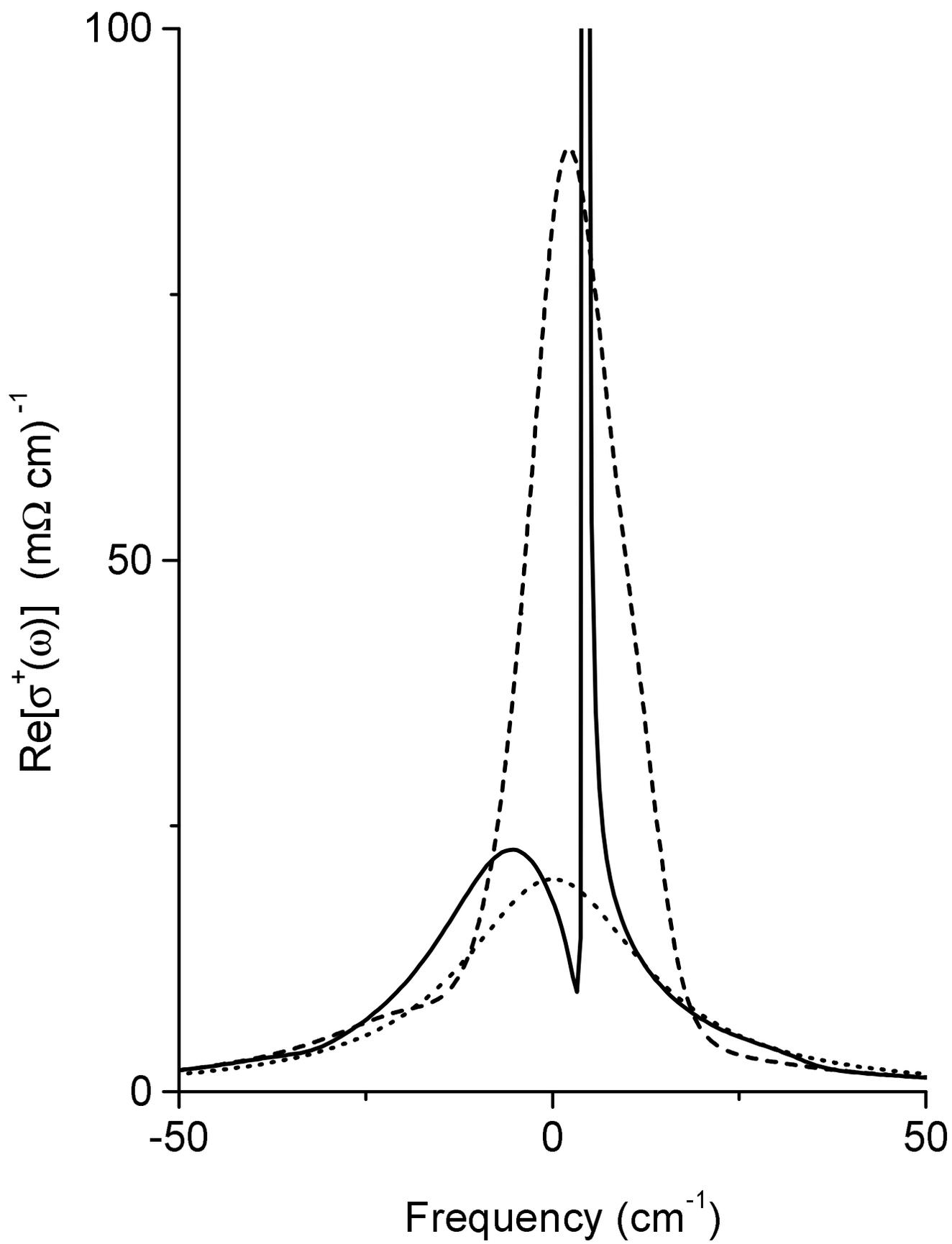
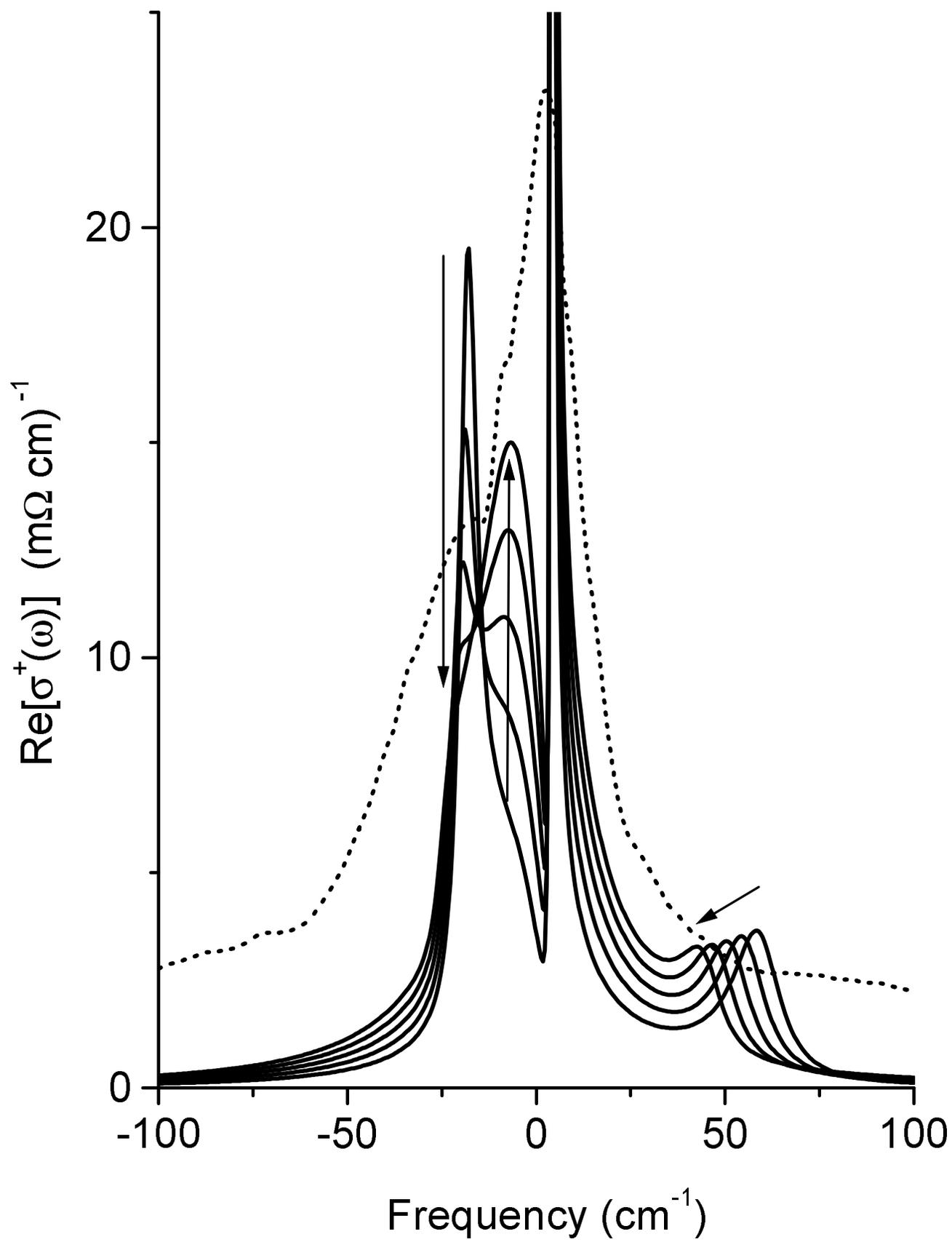
\end{document}